\documentclass[twocolumn,aps,amsmath,amsfonts]{revtex4}
\usepackage{amsthm}
\usepackage{amssymb}
\usepackage{hyperref}

\newcommand{\lp}{\left}
\newcommand{\rp}{\right}
\newcommand{\be}{\begin{eqnarray}}
\newcommand{\ee}{\end{eqnarray}}
\newcommand{\beq}{\begin{equation}}
\newcommand{\eeq}{\end{equation}}
\newcommand{\ba}{\begin{array}}
\newcommand{\ea}{\end{array}}

\newcommand{\braket}[2]{{\langle{#1}|{#2}\rangle}}
\newcommand{\ket}[1]{{|{#1}\rangle}}
\newcommand{\bra}[1]{{\langle{#1}|}}

\newcommand{\mypmatrix}[1]{\begin{pmatrix}#1\end{pmatrix}}
\newcommand{\R}{\mathbb{R}}

\newcommand{\Z}{\mathbb{Z}}

\DeclareMathOperator{\Tr}{Tr}

\DeclareMathOperator{\Imag}{Im}
\DeclareMathOperator{\vspan}{span}

\newcommand{\XOR}{\textsc{xor} }
\newcommand{\XORns}{\textsc{xor}}

\newcommand{\A}{\mathcal{A}}
\newcommand{\B}{\mathcal{B}}
\newcommand{\M}{\mathcal{M}}

\newtheorem{theorem}{Theorem}

\newtheorem{lemma}[theorem]{Lemma}

\begin{document}

\title{Hamiltonian Oracles}
\author{Carlos Mochon}
\email{cmochon@perimeterinstitute.ca}
\affiliation{Perimeter Institute,
31 Caroline St N, Waterloo, ON  N2L 2Y5, Canada}
\date{April 14, 2007}

\begin{abstract}
Hamiltonian oracles are the continuum limit of the standard unitary quantum
oracles. In this limit, the problem of finding the optimal query algorithm
can be mapped into the problem of finding shortest paths on a manifold. The
study of these shortest paths leads to lower bounds of the original unitary
oracle problem. A number of example Hamiltonian oracles are studied in this
paper, including oracle interrogation and the problem of computing the \XOR
of the hidden bits. Both of these problems are related to the study of
geodesics on spheres with non-round metrics. For the case of two hidden
bits a complete description of the geodesics is given. For $n$ hidden bits
a simple lower bound is proven that shows the problems require a query time
proportional to $n$, even in the continuum limit. Finally, the problem of
continuous Grover search is reexamined leading to a modest improvement to
the protocol of Farhi and Gutmann.
\end{abstract}
\maketitle

\section{Introduction}

As a physical theory quantum mechanics distinguishes itself from its
classical counterpart by discretizing certain quantities that were
previously considered continuous. Ironically, it is classical computation
that is inherently a discrete problem, whereas quantum computation involves
a continuous evolution of the state. Nevertheless, when both computational
models are extended to include oracles, the queries are introduced as
discrete events. While there are good physical reasons for such an oracle
model, one is tempted to ask what would constitute half a query to an
oracle. For a standard quantum oracle that applies phases of $1$ and $-1$,
half an oracle call could be an application of the phases $1$ and $i$
respectively. Surely, two calls to the second oracle are at least as
powerful as one call to the standard oracle. Continuing along these lines,
one could envision a fraction $\Delta$ of an oracle query that applies
phases $1$ and $e^{i\pi \Delta}$ respectively. If $1/\Delta$ is an integer,
then that many calls to this new oracle would be at least as good as one
standard oracle query.

Taking the limit $\Delta\rightarrow 0$ one arrives at the Hamiltonian
oracle model, first described by Farhi and Gutmann \cite{FG98}. Roughly
speaking, the evolution in this model is given by the Schr\"odinger
equation
\be
\frac{d}{dt}\ket{\psi(t)} = -i \lp( H_j + H'(t)\rp) \ket{\psi(t)},
\ee
\noindent
where $H_j$ is the Hamiltonian oracle that depends on some hidden
parameter $j$, and $H'(t)$ is a time dependent Hamiltonian that can be
chosen arbitrarily but independently of $j$. The goal of the problem is to
evolve from some fixed initial state to a state that contains some
information of the hidden parameter $j$.

The ability to control $H'$ is equivalent to setting $H'=0$ but being able
to apply fast unitaries as often as one wishes. The standard oracle model
is simply the restriction that unitaries can be applied only at discrete
time intervals. Therefore, from the perspective of the time evolution of
the Gram matrix, continuous oracle algorithms appear as smooth curves
whereas algorithms for the equivalent discrete oracles are piece-wise
continuous approximations of such curves. The question we ask here is:
wouldn't it be easier to study the smooth curves?

This question is in the spirit of recent work by Nielsen et
al.~\cite{Nie05,MDG05} which proposes a similar approach to the study of
quantum circuit lower bounds. The idea is that differential equations are
often easier to solve than difference equations, and that many problems
become simpler in the continuum limit.

Given that Hamiltonian oracles can be obtained as the limit
$\Delta\rightarrow 0$ of discrete oracles, most of the standard techniques
\cite{Amb02,BBC+01,BSS03} for studying discrete oracles can be used to
obtain bounds on Hamiltonian oracles. However, our goal is not to import
bounds from the discrete case into the continuous case.  Rather we seek to
solve Hamiltonian oracle problems by new methods that are intrinsically
continuous, and in some cases geometric, and then export these results back
into the realm of discrete oracle problems in order to prove new lower
bounds. So long as the normalization of the Hamiltonian is chosen so that
the continuous oracle equals $O=e^{-i H}$, the minimum query time needed to
solve the continuous case will be a lower bound on the number of oracle
calls needed to solve the discrete case.

Of course, it is not a priori clear that the continuum limit offers any
simplifications. The reductions of both Ref.~\cite{Nie05} and the present
paper map difficult computer science problems into the problem of finding
shortest curves on manifolds, which is also considered a difficult problem.
Unfortunately, no new bounds on discrete oracles will be obtained in this
paper. Rather, we shall examine Hamiltonian oracles for three problems that
have been solved in the discrete case: Oracle interrogation, the problem of
computing \XORns, and one-item Grover search. However, these examples will
serve as an illustration of both the potential that Hamiltonian oracles
present and some of the techniques that can be used to exploit them.

Oracle interrogation is the problem where the hidden parameter is an
$n$-bit string which can be queried one bit at a time, and we wish to
determine the full hidden string. Though $n$ queries are required to solve
the problem exactly in the discrete setting, van~Dam \cite{vD98} proved
that $n/2 + O(\sqrt{n})$ queries are sufficient to guess the string with
very high probability.

In Section~\ref{sec:oi} we show that the continuous version of oracle
interrogation can be reduced to the study of shortest curves on $S^n$ with a
special metric. For $n=2$ the metric in polar coordinates can be written as
\be
ds^2 = \frac{4}{\pi^2}\lp( d\theta^2 + \tan^2 \theta d\phi^2 \rp),
\ee
\noindent
and the minimum query time needed to solve the problem exactly is equal to
the shortest distance between the points with $\theta=\pi/2$, $\phi=0$ and
$\theta=\pi/4$, $\phi=\pi/4$. We show that a complete set of geodesics can
be constructed for this metric, and that the minimum query time for zero
error is about $90\%$ of the time required to query both bits
separately. This is in contrast with the discrete case where exact
solutions never allow any speedup.

For $n>2$ the metric on $S^n$ is no longer Riemannian but rather of the
more general Finsler type. Though the minimal length curves will not be
constructed for these cases, we prove in Section~\ref{sec:low} 
a simple lower bound on the query time
\be
T \geq \frac{n}{\pi e} + \Omega(1) \simeq 0.117 n
\ee
which applies to the \XOR problem (and hence also to oracle interrogation)
in the bounded error setting. This is an important bound, for if the
Hamiltonian model were significantly faster than the discrete model at
solving oracle interrogation then it would likely be useless for proving
good lower bounds.

The one-item Grover search oracle is studied in Section~\ref{sec:grov}. It
is a simple enough problem that we can observe the transition from the
discrete to the continuum limit. Given the ability to apply unitaries only
at intervals of length $\Delta$, the problem can be solved exactly in a
query time of 
\be
T = \Delta \lp \lceil \frac{\arccos\frac{1}{\sqrt{N}}}
{\arcsin \frac{2\sin(\pi\Delta/2)\sqrt{N-1}}{N} } \rp\rceil
\ee
which implies that for large $N$ we obtain a speedup by a factor of
$\Delta/\sin (\pi \Delta/2)$ relative to the standard discrete oracle
$\Delta=1$. The result can be extended to fixed error, and in every case
half of the above time is needed to solve the problem with probability
greater than one half.

In the continuum limit we obtain a query time
\beq
T = \frac{1}{\pi}\frac{N}{\sqrt{N-1}} \arccos \frac{1}{\sqrt{N}} 
\simeq \frac{1}{2}\sqrt{N}-\frac{1}{\pi}+O\lp(\frac{1}{\sqrt{N}}\rp)
\eeq
for an exact solution, which we prove optimal. The above solution is
shorter asymptotically by an additive constant of $-1/\pi$ than the one
found by Farhi and Gutmann \cite{FG98}. Though the difference is irrelevant
from a practical perspective, the nature of the different solutions is
interesting, and is discussed in Section~\ref{sec:FG}. A similar
improvement for the case $N=2$ was found in Ref.~\cite{CPR99}.

The analysis technique used in this paper is presented in
Section~\ref{sec:sym} and is a variant on the adversary and semidefinite
programing approaches \cite{Amb02,BSS03}, where we study the evolution of
the Gram matrix and use symmetrization to simplify the problem. The
problems considered herein are sufficiently symmetric that this technique
works well. It also has the benefit that it allows the continuous and
discrete problems to be studied together using the same notation. The
divergence between the two formulations can be delayed until the last step
where we consider the dynamics of the Gram matrix.

We note that it does not appear that there is a unique canonical
Hamiltonian for a given unitary oracle. In Section~\ref{sec:can} we show a
pair of unitary oracles that are computationally equivalent, but lead to
different Hamiltonian oracles under the process of replacing ones with $0$
and minus ones with $\pi$. This process also has the undesired effect that
it breaks complex conjugation symmetry. An alternative way of obtaining a
Hamiltonian is to double the query space and replace a one eigenvalue by
two $0$'s and a minus one eigenvalue by $\pi$ and $-\pi$. This essentially
introduces an ``arrow of time'' qubit that allows the choice between the
canonical evolution and its complex conjugate. In fact, it is this form of
oracle that is analyzed in the oracle interrogation and \XOR problems as we
want to ensure that the lower bounds apply to the most general case. For
the Grover search problem the standard oracle was used.

The philosophy that has been adopted in this paper is that Hamiltonian
oracles are a tool in the study of the standard discrete oracles. Being
able to identify a Hamiltonian (possibly from a given set of Hamiltonians)
is also an important problem in experimental physics where the hidden
parameter is some physical constant which we are interested in measuring
\cite{CPR99}. However, the two problems are somewhat different. The
Hamiltonians that correspond to standard unitary oracles typically couple
$O(\log n)$ qubits, where $n$ is the number of possible different
queries. Such couplings are generally not found in nature. Furthermore, the
computational version of the problem only concerns itself with one
resource: query time. In the experimental version of the problem one may
also need to place bounds on the maximum energy of the control Hamiltonian,
how quickly it can be changed and what complexity can be
achieved. Balancing these competing resources, however, is beyond the scope
of this paper.

In the end, all the Hamiltonian oracles studied in this paper were
equivalent up to a constant factor to their discrete counterpart. It is
unclear if such a relationship holds in general and if so, how large can
this constant be? Future work will have to address this question along the
road to finding new lower bounds from Hamiltonian oracles.

\subsection*{Prior work}

The first paper to study Hamiltonian oracles from a quantum computation
perspective is the work of Farhi and Gutmann \cite{FG98} as discussed
above.

The paper by Fenner \cite{Fen00} reexamines the continuous Grover
search with a goal of finding a Hamiltonian that matches the discrete
case step by step. However, in their construction they allow a total
Hamiltonian that is a commutator of the oracle and control Hamiltonians,
whose physical motivation is unclear.

The paper by Roland and Cerf \cite{RC03} also compares the discrete and
continuous version of Grover search and studies the simulation of the
continuous algorithm by a discrete quantum computer.

Most of the subsequent work involving Hamiltonian oracles studied the
problem of spatial search \cite{CG03,CG04}, which is a variant of Grover
search where the database has some spatial arrangement and only local moves
are permitted. The algorithms for these problems employ the continuous
quantum walk \cite{FG97}.

There are also many papers that study the problem of identifying a
Hamiltonian, though their goals are generally different from ours. For
instance, Ref.~\cite{AMP01} studies the time-energy uncertainty relation as
applied to Hamiltonian identification whereas Ref.~\cite{JB01} shows that
in principle a set of Hamiltonians can be distinguished, though the
efficiency is not considered. The relation between Hamiltonian oracles and
identifying Hamiltonians in the laboratory was also discussed above:
Childs, Preskill and Renes \cite{CPR99} continued the work on Hamiltonian
oracles, with a view towards exporting the knowledge of quantum computation
to the realm of experimental physics. In the same spirit is the work of
quantum parameter estimation for dynamical systems, such as the paper by
Mabuchi \cite{Mab96}.

Finally, the study of Hamiltonian oracles can be recast into a number of
formalisms including time optimal control \cite{KBG00} which also greatly
benefits from geometric approaches. As a Hamiltonian oracle problem can be
studied as a single bipartite Hamiltonian where one can perform arbitrary
operations on one side only, this can be translated into the language of
optimal control by identifying the oracle Hamiltonian as a drift
Hamiltonian, and the subgroup of allowed operations $K\subset G$ as those
that act on only one subsystem. Once again, however, the typical
Hamiltonians that are of interest in one field are fairly different from
those of the other.

\section{\label{sec:sym}Models and Methods}

Below we shall introduce a more formal definition of the Hamiltonian oracle
model, which will be presented in a language that emphasizes its
connections to discrete oracles. We shall use the description of oracle
problems as an Alice-Bob game, with Alice taking the place of the
oracle. This will facilitate the translation of the problem into a
semidefinite program using Kitaev's construction for coin-flipping
\cite{Kitaev}. The resulting semidefinite program will be equivalent to the
one of Barnum, Saks and Szegedy \cite{BSS03}, though it will be easier to
symmetrize. Most of the discussion in this section has appeared elsewhere
and is intended mainly for review purposes and to fix the notation used for
the rest of the paper.

In the Alice-Bob game description of the oracle problem Alice starts with a
hidden string (or superposition of strings) in a Hilbert space $\A$. Bob
can query Alice by sending a message in some space $\M$. Alice always
applies some known fixed unitary (or Hamiltonian) to $\A\otimes\M$ and
returns $\M$ to Bob. Of course, Bob is allowed to have his own private
Hilbert space $\B$, however it will never be explicitly referenced
as everything will be described from Alice's perspective.

In the end Bob must guess some property of the hidden string, and send his
guess to Alice in space $\M'$, who determines whether it is correct or
not. We say that Bob wins when Alice accepts his answer, and the goal is to
maximize this probability.

In principle, given a strategy for Bob, we need to try it against each of
the possible hidden strings, one at a time. Because we are interested in
the worse-case success probability, we take the minimum over the success
probabilities for the different possible hidden strings. However, as is
common in adversary methods, Alice can start with a superposition over
different possible input strings. In such a case the worse-case success
probability can be calculated by a single run through the Alice-Bob game.
However, the operation which computes this final worse-case success
probability is not a physical quantum measurement but rather just a
linear expression involving Alice's final density operator. Nevertheless,
it will be Bob's goal to use his interactions with Alice in order to attain
a final density operator that maximizes the expression for the success
probability. We shall say more about this final operation below.

Formally, we define an oracle problem by three Hilbert spaces $\A$, $\M$
and $\M'$, together with an initial pure state $\ket{\psi_0}$ on $\A$, a
unitary operator $O$ (or Hermitian operator $H$ in the continuous case) on
$\A\otimes\M$, and a set of positive operators $\{\Pi_x\}$ on
$\A\otimes\M'$ labeled by an index $x$ which usually ranges over the set of
hidden strings. For the discrete case we also specify a positive number
$\Delta$ corresponding to an interval of time.

A protocol for an oracle problem is given by a positive time $T$ (divisible
by $\Delta$ in the discrete case), a success probability $P_{win}$, a pair
of functions $\rho(t)$ and $\tilde \rho(t)$ for $0\leq t \leq T$ (valued at
integer multiples of $\Delta$ for the discrete case) and a final matrix
$\tilde \rho'$. We require that $\rho(t)$, $\tilde \rho(t)$ and $\tilde
\rho'$ be positive operators on the spaces $\A$, $\A\otimes\M$ and
$\A\otimes\M'$ respectively. They must satisfy the following equations:

\begin{itemize}
\item Initialization:
\be
\rho(0) &=& \ket{\psi_0}\bra{\psi_0}_\A.
\ee
\item Bob's action (for $0\leq t \leq T$):
\be
\Tr_\M[\tilde \rho(t)]&=&\rho(t).
\label{eq:rt}
\ee
\item Alice's action
\begin{itemize}
\item for discrete time ($0\leq t \leq T-\Delta$)
\be
\rho(t+\Delta) &=& \Tr_\M[O \tilde \rho(t) O^{-1}].
\label{eq:O}
\ee
\item for continuous time ($0\leq t\leq T$)
\be
\frac{d}{dt}\rho(t) &=& -i \Tr_\M[H  \tilde \rho(t) - \tilde \rho(t) H].
\label{eq:H}
\ee
\end{itemize}
\item Bob's output:
\be
\Tr_{\M'}[\tilde \rho']&=& \rho(T).
\ee
\item Answer verification (for every $x$):
\be
P_{win} \leq \Tr[ \Pi_x \tilde\rho' ].
\label{eq:pwin}
\ee
\end{itemize}

A standard discrete-time oracle problem will have $\Delta=1$, however, the
above formulation allows us to pass to the continuous time limit by
defining $O=e^{-i\Delta H}$ and then taking the limit $\Delta\rightarrow
0$.

The basic goal of the problem is to choose the protocol $\rho(t)$, $\tilde
\rho(t)$ and $\tilde \rho'$ as to maximize the probability of winning
$P_{win}$, for a given time $T$. Of course, eventually one wants to invert
the relation: fix $P_{win}$ and find the smallest $T$ for which it can be
achieved as a function of some scaling of the problem.

The above formulation should be understood as follows: Say Alice has a
density operator $\rho(t)$ on $\A$ at a given time $t$. When Bob queries
Alice by sending a message in the space $\M$, Alice ends up with a density
operator $\tilde \rho(t)$ on the larger space $\A\otimes\M$. This operator
must satisfy the consistency condition given by Eq.~(\ref{eq:rt}) because
Bob cannot affect the state of $\A$. Having received Bob's message, Alice
applies the oracle operation and returns $\M$ to Bob, ending up with a new
state defined by Eq.~(\ref{eq:O}) or Eq.~(\ref{eq:H}).

To relate the above definition to the standard oracle model we let $\A$ be
the Hilbert space spanned by the set of hidden strings. Then $O = \sum_{x}
\ket{x}\bra{x}_\A\otimes O_x$ where $O_x$ are the standard oracle operators
on $\M$ given hidden parameter $x$. For the continuous case we similarly
have $H = \sum_{x} \ket{x}\bra{x}_\A\otimes H_x$.

A good guess for the final operation would be the two-outcome POVM
$\{\Pi,I-\Pi\}$, where $\Pi = \sum_x \ket{x}\bra{x}_\A
\otimes\ket{f(x)}\bra{f(x)}_{\M'}$ and $f(x)$ is the target function 
to compute such as \XORns. We would then declare Bob a winner only if the
first outcome was obtained, thereby setting $P_{win} = \Tr[\Pi \tilde
\rho']$. However, this only computes the average success probability rather
than the worse-case success probability. Instead, the correct prescription
is to use Eq.~(\ref{eq:pwin}) with 
$\Pi_x= \ket{x}\bra{x}_\A\otimes\ket{f(x)}\bra{f(x)}_{\M'}/ 
|\braket{x}{\psi_0}|^2$, so that $\Tr[\Pi_x\tilde\rho']$ will be the
probability of Bob correctly answering given that the
hidden string was $x$.

For the discrete oracle case deriving the above semidefinite program is
fairly simple. Clearly no matter what actions Bob performs, Alice's
density operators must satisfy the above equations. On the other hand,
because Alice starts with a pure state and makes no measurements, Bob can
keep the purification of Alice's state and therefore force any evolution
consistent with the above equations.

To arrive at the continuous case simply define $\Delta=2^{-k}$ and
$O=e^{-i\Delta H}$ for $k\in\Z^+$. Given some fixed $T$, let $P_{win}(k)$
be the maximum over all protocols for the discrete oracle problem with a
given $k$. Trivially, $P_{win}(k+1)\geq P_{win}(k)$. If we also defined the
problem so that $P_{win}(k)\leq 1$ for all $k$, then in the limit
$k\rightarrow \infty$ we converge to a well defined $P_{win}(\infty)$. This
can be taken as a formal definition of the Hamiltonian oracle problem. In
this limit we can replace the discrete evolution with the continuous
evolution given by Eq.~(\ref{eq:H}), so long as we restrict $\tilde
\rho(t)$ to be continuous (or more generally measurable if we use the
integral form of the equation).

The more traditional definition of a Hamiltonian oracle is that there is a
set of Hamiltonians $\{H_x\}$ acting on a space $\M$. Bob can append a set
of extra qubits with the space $\B$ on which the Hamiltonians act
trivially. He can then control the system by either adding an extra
Hamiltonian $H'$, by periodically applying unitary operators, or by
conjugating $H_x\otimes I_\B$ by some time dependent unitary of his
choosing, so long as these operations don't depend on the hidden parameter
$x$. These three variants are all equivalent, and equivalent to the model
where all three activities can be done simultaneously. Though proving the
equivalence of these models is beyond the scope of this paper, it is not
hard to see that the above continuous SDP serves as a lower bound for all
the models, again by the argument that no matter what Bob does, the qubits
in Alice's possession are restricted to evolve according to the above
equations.

We note in closing that the above SDP can be separated into two problems:
The first is finding the set of attainable final density operators
$\rho(T)$. The second problem involves finding the optimal $\tilde
\rho'$ and maximal $P_{win}$ given $\rho(T)$. Solving the second problem
is often easy, leading to a function $P_{win}(\rho(T))$. Therefore, most of
the effort below will involve searching for the evolution towards a good
final density operator $\rho(T)$.

\subsection{\label{sec:can}On canonical Hamiltonians}

In this section we shall examine an oddity that arises in the transition
from discrete to continuous oracles. Clearly given an oracle unitary, $O$,
there are infinitely many Hamiltonians $H$ such that $O=e^{-i H}$. However,
if $O$ is a standard oracle with only eigenvalues $1$ and $-1$ then a
canonical Hamiltonian can be defined by the process of replacing the one
eigenvalues by $0$ and the minus one eigenvalues by $\pi$.

Unfortunately, while the above mapping does associate a unique Hamiltonian
to each unitary oracle, there are cases when unitary oracles of equivalent
computational power are mapped into Hamiltonians of different
computational power.

Consider for instance the oracle with a hidden bit $b\in\{0,1\}$, and the
two unitary oracles
\be
O_0 = \mypmatrix{1 & 0 & 0 \cr 0 & 1 & 0 \cr 0 & 0 & 1},\qquad
O_1 = \mypmatrix{1 & 0 & 0 \cr 0 & -1 & 0 \cr 0 & 0 & -1},
\ee
\noindent
where the complete oracle in the notation of the previous section would be
$O=\ket{0}\bra{0}_\A \otimes O_0 + \ket{1}\bra{1}_\A \otimes O_1$. A
different pair of oracles for the same problem are given by 
\be
O_0' = \mypmatrix{1 & 0 & 0 \cr 0 & 1 & 0 \cr 0 & 0 & -1},\qquad
O_1' = \mypmatrix{1 & 0 & 0 \cr 0 & -1 & 0 \cr 0 & 0 & 1}
\ee
\noindent
with $O'$ defined similarly. The two pairs of oracles are clearly
equivalent, as one can simulate one with the other by simply applying a
phase flip to the third basis state.

Now consider the Hamiltonians obtained from the above unitaries by the
standard eigenvalue replacement
\be
H_0 = \pi \mypmatrix{0 & 0 & 0 \cr 0 & 0 & 0 \cr 0 & 0 & 0}, \qquad
H_1 = \pi \mypmatrix{0 & 0 & 0 \cr 0 & 1 & 0 \cr 0 & 0 & 1}
\ee
\noindent
and
\be
H_0' = \pi \mypmatrix{0 & 0 & 0 \cr 0 & 0 & 0 \cr 0 & 0 & 1},\qquad
H_1' = \pi \mypmatrix{0 & 0 & 0 \cr 0 & 1 & 0 \cr 0 & 0 & 0},
\ee
where again $H=\ket{0}\bra{0}_\A \otimes H_0 + \ket{1}\bra{1}_\A \otimes
H_1$ and similarly for $H'$. With the first oracle pair it takes one unit
of time to perfectly distinguish between the two Hamiltonians, whereas with
the second pair only half a unit of time is required.

In general, the existence of many choices for the Hamiltonian oracle is not
a problem. So long as the normalization is chosen so that $e^{-i H}$ is
computationally equivalent to the discrete oracle that needs lower
bounding, one may choose any Hamiltonian that is easy to study.

\subsection{Reduction by preponderance of symmetry}

In the next two sections we shall show how oracle problems with large
amounts of symmetry can be simplified because they always have an optimal
solution that shares the symmetry of the problem, and therefore the search
for the optimal solution can be conducted over the smaller space of density
operators that are invariant under the action of the symmetry group.

We say that a group $G$ is compatible with an oracle problem
$\ket{\psi_0}$, $O$ (or $H$ for the continuous case), and $\{\Pi_x\}$ if
there exists unitary representations $R_\A$ on $\A$, $R_\M$ on $\M$ and
$R_{\M'}$ on $\M'$ such that for all $g\in G$ we have
\be
R_\A(g) \ket{\psi_0}
&=& \ket{\psi_0},\\
R(g) O R(g^{-1}) &=& O,\\
R(g) H R(g^{-1}) &=& H,\\
\{R'(g) \Pi_x R'(g^{-1})\}
&=& \{\Pi_x\},
\ee
\noindent
where for simplicity we have introduced $R(g)=R_\A(g)\otimes R_\M(g)$ and
$R'(g)=R_\A(g)\otimes R_{\M'}(g)$. In the last equation, the symmetry does
not need to leave the operators $\Pi_x$ element-wise invariant, but must
leave the set invariant.

Given a solution $\tilde \rho(t)$ to a $G$ compatible oracle problem, 
we can define for $g\in G$
\be
\tilde \rho_g(t) =
R(g) \tilde \rho(t) R(g^{-1})
\ee
\noindent
for $t\in[0,T]$ and similarly
\be
\tilde \rho_g' =
R'(g) \tilde \rho' R'(g^{-1}),
\ee
\noindent
which will be solutions to the oracle problem with the same success
probability (as can easily be verified). Naturally, the reduced density
operator on Alice's side will be $\rho_g(t) = 
\Tr[\tilde\rho_g(t)]= R_\A(g) \rho(t) R_A(g^{-1})$.

Because the equations are all linear, we can also take linear combinations
of solutions. Given a solution $\tilde \rho(t)$ define
\be
\tilde \rho_G(t) = \frac{1}{|G|} \sum_{g\in G} \tilde \rho_g(t)
\ee
\noindent
and similarly for $\tilde \rho_G'$. These must also be a solution of the
oracle problem with the same success probability as $\tilde \rho(t)$ and
$\tilde \rho'$. Note that this is a non-vanishing solution because $\tilde
\rho(t)=0$ cannot be a solution of the equations. In fact, the equations
impose conservation of trace so that for all $t$ we have
$\Tr[\tilde\rho_G(t)]=\braket{\psi_0}{\psi_0}$.

We call a solution $G$-invariant if for all $g\in G$ and $t\in[0,T]$ it
satisfies $R(g)\tilde \rho(t) R(g^{-1})=\tilde \rho(t)$, and furthermore
$R'(g)\tilde \rho' R'(g^{-1})=\tilde \rho'$ again for all $g\in G$. A $G$
invariant solution will also imply that $R_\A(g)\rho(t)R_\A(g^{-1}) =
\rho(t)$ for all $g\in G$ and at all times.

It is not hard to see that given any solution $\tilde \rho(t)$,
the solution $\tilde \rho_G(t)$ constructed from it as above will be
$G$-invariant, and will have the same probability of success. We have
therefore proven the following lemma:

\begin{lemma}
Given an oracle problem that is compatible with a group $G$, the set of
success probabilities $P_{win}$ that can be achieved with query time $T$
will not be altered if we restrict the space of solutions to those that are
$G$-invariant.
\end{lemma}

The lemma allows us to concentrate only on $G$-invariant solutions when
studying both upper and lower bounds.

\subsection{\label{sec:sigma}Further reductions for standard oracles}

Up to this point the set of allowed oracle unitaries or Hamiltonians has
been left unrestricted, but now we shall focus on the standard oracles that
change the phase of the states in $\M$ based on the value of the hidden
oracle string.

Given a basis $\{\ket{j}_\A\}$ for $\A$ and a basis $\{\ket{k}_\M\}$ for
$\M$, which we refer to as the computational bases, we say that $O$ or $H$
is in standard form if it can be written as
\be
\sum_{j,k} C_{j,k} \ket{j}\bra{j}_\A\otimes \ket{k}\bra{k}_\M,
\ee
\noindent
where the coefficients $C_{j,k}$ are real numbers in the case of a
Hamiltonian oracle and phases in the case of a unitary oracle.

We further assume that the oracle problem is compatible with a group $G$
and that the action of this group on the space $\M$ is by permutation of
the basis:
\be
R_\M(g) \ket{k}_\M = \ket{s(g) k}_\M
\ee
for all $k$ and $g\in G$, where $s(g)$ is a homomorphism from $G$ to the
symmetric group $S_{M}$ and $M=|\M|$. For simplicity we assume here that
the index set $k$ ranges from $1$ to $M$ so that $S_{M}$ acts naturally on
the index set by permutation. In such a case the symmetry group $G$ can be
extended to $G_Q=(\Z_2)^M\rtimes G$, where the semidirect product is defined
by $(x',g')(x,g)=(x'\oplus s(g') x,g' g)$ with $x\in(\Z_2)^M$ and $M$-digit
binary string and $s(g')\in S_M$ acting on it by permutation.

We can define representations of $G_Q$ on the spaces $\A$ and $\M'$ by
$R_\A(x,g)=R_\A(g)$ and $R_{\M'}(x,g)=R_{\M'}(g)$. The nontrivial
extension is the representation of $G_Q$ on $\M$ defined by
\be
R_\M(x,g) \ket{k} =  (-1)^{x_{s(g)k}} \ket{s(g) k},
\ee
\noindent
where $x_k$ denotes the $k$th bit of $x$. It is simple to verify that if
the a standard oracle problem is compatible with $G$, then it will also be
compatible with $G_Q$. 

A $G_Q$-invariant operator $\tilde\rho$ on $\A\otimes\M$ must have the
block diagonal form
\be
\sum_k \sigma_k \otimes \ket{k}\bra{k}_\M,
\ee
\noindent
where the $\sigma_k$ are positive operators on $\A$. The block
diagonalization follows simply by considering the action of group elements
of the form $(x,1)$, where $x$ is a string with a single $1$ entry.

The $\sigma_k$ are further restricted as follows:
\begin{itemize}
\item The operators $\sigma_k$ must be invariant under the stabilizer of
$k$, that is, for all $g\in G$ such that $s(g)k=k$ we must have
$R_\A(g)\sigma_k R_\A(g^{-1}) =\sigma_k$.
\item If there exists a $g$ such that $s(g)k=k'$ then $\sigma_{k'}=R_\A(g)
\sigma_k R_\A(g^{-1})$ for any such $g$.
\end{itemize}

The condition  $s(g)k=k'$ for some $g\in G$ defines an equivalence
relationship on the integers $[M]$, and allows us to divide them
into equivalence classes. The most general $G$-invariant $\tilde \rho$ can
therefore be specified by only one matrix $\sigma_k$ for each equivalence
class, which must be invariant under the stabilizer of $k$.

Of course, all of the above discussion would be moot if we had just begun
with the symmetry group $G_Q$ in the first place. However, in this paper
we shall choose $G$ to correspond with the symmetry of the classical
problem, and then $G_Q$ will be the extended symmetry that appears in the
quantum case.

We conclude this section by noting how the symmetry simplifies density
operators $\rho$ on $\A$ (as opposed to $\tilde \rho$ on $\A\otimes\M$ as
we have been discussing thus far). If the decomposition of the
representation $R_\A$ into irreducible representations contains at most one
copy of each irrep, then by Schur's lemma the most general $G$-invariant
$\rho$ has the form
\be
\sum_{\alpha}  a_\alpha^2 P_\alpha,
\ee
\noindent
where $\alpha$ ranges over the irreps appearing in $R_{\A}$, $P_\alpha$ is
the projector onto the irrep, and the $a_\alpha^2$ are non-negative
constants that sum to $\braket{\psi_0}{\psi_0}$ which we assume here is
one. It will be convenient to deal with the vectors $(a_0,a_1,\dots)$ which
specify a point on the unit sphere for some dimension. This point on the
sphere will be a complete description of the state of the protocol at a
given instant of time, or equivalently, of Bob's knowledge at that instant
of time.

\section{\label{sec:grov}One-item Grover search}

Here we study the Grover search problem under the promise that exactly one
item is marked. The goal, as usual, is to identify the marked item. Though
this problem has been extensively studied in the literature, it will
provide a good example for the ideas discussed in the previous section, and
as a comparison of the discrete and continuous oracle models. Furthermore,
we shall find a modest improvement to the protocol found by Farhi and
Gutmann \cite{FG98}.

Our strategy below, after defining the problem in the new notation, will be
to identify the symmetry group of the problem and use it to reduce the
search space of potential solutions. Within this reduced space we will then
identify the initial state and the set of final states from which the
marked state can be identified with small error. Finally, we study the
dynamics needed to evolve from the initial state to these good final
states, which will tell us the query time needed to solve the oracle
problem. Note that it is only in this last step that the discrete and
continuous oracle models need to be handled separately.

\subsection{Problem definition}

Fix an integer $N>1$ and define
\be
\A &=& \vspan\{\ket{j}\text{ for }j\in[N]\},\\
\M=\M' &=& \vspan\{\ket{j}\text{ for }j\in\{0\}\cup[N]\},
\ee
and on these spaces we define
\be
\ket{\psi_0} &=& \ket{+}_\A 
\equiv \frac{1}{\sqrt{N}}\sum_{j=1}^{N} \ket{j}_\A,\\
H &=& \pi \sum_{j=1}^N \ket{j}\bra{j}_\A \otimes \ket{j}\bra{j}_\M,\\
O_\Delta &=& I - (1-e^{-i\pi\Delta}) 
\sum_{j=1}^N \ket{j}\bra{j}_\A \otimes \ket{j}\bra{j}_\M,\\
\Pi_j &=& N \ket{j}\bra{j}_\A \otimes \ket{j}\bra{j}_{\M'}\qquad
\text{for $j\in[N]$}.
\ee

\noindent
Note that the normalization of $H$ is chosen so that
$O_\Delta=e^{-i\Delta H}$, and the unit of time is chosen so that
$\Delta=1$ corresponds to the standard discrete time oracle. In both cases,
the query space $H$ has a state $\ket{0}$ which is left invariant by $H$
and $O$. This is the null query.

The normalization of $\ket{\psi_0}$ to a unit vector, though natural from a
quantum mechanical perspective, will imply that Alice's reduced density
operator will not be the Gram matrix but rather the Gram matrix scaled by
$1/N$. This is also the source of the factor of $N$ in the operators
$\Pi_j$. After the symmetrization below we will be able to replace the $N$
projectors $\{\Pi_j\}$ with the single projection operator $\Pi=\sum_j
\Pi_j/N$ because the worse-case and average-case success probabilities will
be equal.

\subsection{Symmetrization}

The natural symmetry group for the problem is $G=S_N$ which acts by
permutation on $\A$ and on the last $N$ states of $\M$, but leaves
$\ket{0}_\M$ invariant. With these definitions the oracle is compatible
with the symmetry group.

The most general density matrix $\rho$ on $\A$ which is $G$-invariant is
given by
\be
\rho = x \ket{+}\bra{+} + y \lp( I -
\ket{+}\bra{+} \rp).
\ee
\noindent
The two parameters are related because we require
$x+(N-1)y=\Tr[\rho]=\braket{\psi_0}{\psi_0}=1$. Therefore $\rho$ (and hence
the state of the system at any given time) depends on the single parameter
$x$.

We now turn to the symmetrization of operators $\tilde \rho$ on
$\A\otimes\M$. As the oracle is of standard form, and the symmetry group
acts by permutation on $\M$, we can apply the results of
Section~\ref{sec:sigma}. The most general $\tilde \rho$ consistent with the
symmetries of the problem has the form
\beq
\tilde \rho = 
\sigma_0 \otimes \ket{0}\bra{0}_\M +
\sum_{j=1}^{N} R_\A(g_{1,j})\sigma_1 R_A(g_{1,j}^{-1})
\otimes\ket{j}\bra{j}_\M,
\eeq
\noindent
where $g_{1,j}$ is any element that maps $1$ to $j$. 

The matrix $\sigma_0$ must be invariant under
the complete $S_N$ so that its most general form is
\be
\sigma_0 = x_0 \ket{+}\bra{+}+y_0(I-\ket{+}\bra{+}),
\ee
\noindent
where positivity demands $x_0\geq 0$ and $y_0\geq 0$.

On the other hand, the matrix $\sigma_1$ need be invariant only under the
subgroup $S_{N-1}$ that leaves $\ket{1}_\M$ invariant. Under the full $S_N$
we saw that $\A$ decomposes into two irreps: the space spanned by
$\ket{+}_\A$ and its orthogonal complement. Under the restriction to
$S_{N-1}$, the first representation will naturally still be irreducible,
but the second one will decompose into the space spanned by
\be
\ket{-_1}=\sqrt{\frac{N-1}{N}}\ket{1}-
\sqrt{\frac{1}{N(N-1)}}\sum_{j=2}^N \ket{j}
\label{eq:ket-j}
\ee
\noindent
and its orthogonal complement, leading to a decomposition of $\A$ into
three irreps. However, since the first two are both the trivial
representation, Schur's lemma does not prevent them from sharing
off-diagonal terms and therefore the most general $S_{N-1}$ invariant
operator on $\A$ has the form 
\beq 
\sigma_1 = \mypmatrix{\ket{+}&\ket{-_1}} \mypmatrix{a & b\cr b^* & c}
\mypmatrix{\bra{+}\cr\bra{-_1}} + d\, P_1^\perp,
\eeq
\noindent
where $P_1^\perp = (I-\ket{+}\bra{+}-\ket{-_1}\bra{-_1})$ is the projector
onto the orthogonal complement of the space that contains $\ket{+}$ and
$\ket{1}$. Positivity of $\sigma_1$ requires $a\geq 0$, $c\geq 0$, $d\geq
0$ and $|b|^2\leq ac$.

To conclude, we compute the partial trace of a given $\tilde \rho$ of the
above form. It is given by the sum of the projections of $\sigma_i$ onto
the invariant subspaces of the full $S_N$:
\be 
\Tr_\M[\tilde \rho] &=& (x_0+ a N) \ket{+}\bra{+} 
\label{eq:Gcons}
\\\nonumber 
&& + (y_0 + c\frac{N}{N-1} 
+ d \frac{N (N-2)}{N-1}) \lp( I - \ket{+}\bra{+}\rp).
\ee

\subsection{Boundary conditions}

We now proceed to treat $\rho$, and consequently $x$, as a function of
time. The initial condition at time $t=0$ is fairly simple
$\rho(0)=\ket{\psi_0}\bra{\psi_0}$ and hence $x(0)=1$.

We need to determine what values for $x(T)$ are acceptable as final
conditions. The final probability of success $P_{win}$ depends only on
$x(T)$. Because of the symmetry of the problem, the optimal measurement is
the pretty-good measurement and the success probability is given
\cite{BKM97,SKI98,Moc05} by
\be
P_{win}(x(T))&=&\frac{1}{N}\lp(\Tr\sqrt{\rho(T)}\rp)^2 \\\nonumber
&=& \frac{1}{N} \lp(\sqrt{x(T)} + \sqrt{(N-1)(1-x(T))}\rp)^2.
\ee
In particular, a zero error outcome requires $x(T)=1/N$. On the other hand
$x(T)=1/2$ implies $P_{win}=1/2+\sqrt{N-1}/N$, so that a solution with some
fixed error $P_{win}>1/2$ as $N\rightarrow\infty$ requires at a minimum
$x(T)<1/2$.

\subsection{Dynamics}

Bob's task is now clear. He must use the dynamics of the system so that $x$
evolves from $1$ at time $t=0$, and decreases as quickly as possible, past
$1/2$ for constant error and stopping at $1/N$ for zero error.

\subsubsection{Continuous time}

We begin with the continuous case, with dynamics given by Eq.~(\ref{eq:H}),
which leads to a differential equation for $x(t)$:
\be
\frac{dx(t)}{dt} &=& -i \bra{+}\Tr_M[H\tilde \rho(t)
-\tilde\rho(t) H]\ket{+}
\nonumber\\
&=& -\frac{i \pi}{\sqrt{N}} \sum_{j=1}^N \bra{j}\sigma_j(t)\ket{+} - 
\bra{+}\sigma_j(t)\ket{j}
\nonumber\\
&=& 2\pi \sqrt{N-1} \Imag[b(t)].
\ee
\noindent
The differential equation for $y(t)$ is uninteresting, as it is related to
the above by the normalization condition $x(t)+(N-1)y(t)=1$.

Bob controls the dynamics via his choice of $\tilde \rho(t)$ at each time,
which he clearly would like to choose so that the imaginary part of $b(t)$
is as negative as possible. By positivity of $\tilde\rho$ we know that
$|b(t)|^2\leq a(t)c(t)$ whereas the constraint
$\Tr_\M[\tilde\rho(t)]=\rho(t)$ translates via Eq.~(\ref{eq:Gcons}) into
$x(t) = x_0(t) + a(t) N$ and $y(t)=(1-x(t))/(N-1)= y_0 + c(t) N/(N-1) + d(t)
N(N-2)/(N-1)$. Combining these constraints we see that
\be
-\Imag[b(t)]\leq \frac{1}{N}\sqrt{x(t) (1- x(t))}
\ee
\noindent
with equality clearly achievable. The evolution of $x(t)$ following the
optimal protocol is therefore given by
\be
\frac{dx(t)}{dt} = - 2\pi \frac{\sqrt{N-1}}{N} \sqrt{x(t) (1-x(t))}
\ee
\noindent
which is solved by
\be
x(t) = \cos^2 \lp( \pi \frac{\sqrt{N-1}}{N} t\rp),
\ee
\noindent
where we have already included the initial condition $x(0)=1$. Continuous
Grover search can therefore be solved exactly in a time
\be
T = \frac{1}{\pi}\frac{N}{\sqrt{N-1}} \arccos \frac{1}{\sqrt{N}} 
\simeq \frac{1}{2}\sqrt{N},
\ee
\noindent
whereas solving with a fixed error greater than one half requires a query
time of at least $N/(4\sqrt{N-1})$.

\subsubsection{Discrete time}

For the discrete case we have
\be
x(t+\Delta) &=& \bra{+}  \Tr_\M[O\tilde\rho(t) O^{-1}] \ket{+}
\\\nonumber
&=& x_0(t) + N
\mypmatrix{\alpha & \beta}
\mypmatrix{a(t) & b(t) \cr b^*(t) & c(t)} \mypmatrix{\alpha^* \cr \beta^*},
\ee
\noindent
where
\be
\alpha &=& \bra{+}O_1\ket{+} = 
1 - \frac{1-e^{-i\pi\Delta}}{N},
\\
\beta &=& \bra{+}O_1\ket{-_1} =
-\frac{(1-e^{-i\pi\Delta})\sqrt{N-1}}{N},
\ee
\noindent
and $O_1 = I-(1-e^{i\pi\Delta})\ket{1}\bra{1}$ is the oracle operator when
query one is issued. Note that $|\alpha|^2+|\beta|^2=1$.

Before solving the general case of the above equation, we must address what
happens in the last query.  Assume that at some time $t$ we have $x(t)>1/N$
but for some setting of the parameters we can achieve $x(t+\Delta)\leq
1/N$.  Because $x(t+\Delta)$ is continuous in the parameters of
$\tilde\rho(t)$, and we could have also issued a null query (i.e., setting
$a(t)=b(t)=c(t)=d(t)=0$, $x_0(t)=x(t)$, and $y_0(t)=y(t)$ so that
$x(t+\Delta)=x(t)$) there must be a query such that $x(t+\Delta)=1/N$, and
therefore the problem can be solved exactly in $t+\Delta$ queries.

For all other times we know that for any choice of $\tilde\rho(t)$ we must
have $x(t+\Delta)> 1/N$. In this case Bob simply wishes to make
$x(t+\Delta)$ as small as possible. Given that any solution with $x_0(t)>0$
we can always find a better solution by choosing $x_0(t)=0$ and $a(t)=
x(t)/N$. The only two constraints in which $a(t)$ or $x_0(t)$ appear are
$x(t) = x_0(t) + N a(t)$ which is satisfied by the new variables, and
$a(t)c(t)\geq|b(t)|^2$ which is also satisfied as we have not decreased
$a(t)$.

Similarly, given any assignment of the above variables, we can always
set
\be
b(t) = -\sqrt{a(t) c(t)} \frac{\alpha^* \beta}{|\alpha \beta|}
\ee
\noindent
which will not increase $x(t+\Delta)$. With these simplifications, the
optimal solutions must be of the form:
\be
x(t+\Delta) = N \lp(|\alpha| \sqrt{\frac{x(t)}{N}} - |\beta|\sqrt{c(t)}\rp)^2
\ee
\noindent
for some $c(t)\in[0,(1-x(t))/N]$. However, since by assumption $x(t+\Delta)$
cannot be zero, it must be minimized by $c(t)=(1-x(t))/N$. We are left with
the recursive relation
\be
x(t+\Delta) = \lp(|\alpha| \sqrt{x(t)} - |\beta|\sqrt{1-x(t)}\rp)^2
\ee
\noindent
which is solved, with starting point x(0)=1, by
\be
x(t) = \cos^2\lp( \arcsin(|\beta|) \frac{t}{\Delta}\rp)
\ee
\noindent
yielding an exact solution in a query time
\beq
T = \Delta \lp \lceil \frac{\arccos\frac{1}{\sqrt{N}}}{\arcsin \frac{
2\sin(\pi\Delta/2)\sqrt{N-1}}{N} 
} \rp\rceil.
\eeq
\noindent
As before a fixed error greater than one half can also be attained in
approximately half the time.

\subsection{\label{sec:FG}Discussion}

Just as in the discrete case, the optimal continuous protocol for
one-item Grover search can be described as a rotation in the
two-dimensional subspace that contains the vectors $\ket{+}$ and the marked
state $\ket{j}$. This rotation is effectuated by the Hamiltonian
\be
H_{total} = H_j + H' = \pi \ket{j}\bra{j} + \pi\frac{N-2}{N}\ket{+}\bra{+},
\ee
\noindent
where $H_j$ is the oracle Hamiltonian if the hidden string is $j$, and $H'$
is the $j$ independent Hamiltonian that defines the algorithm. In the
orthonormal basis $\ket{+}$ and $\ket{-_j}$ for the relevant two
dimensional subspace the above equation reads
\be
H_{total} &=& \frac{\pi}{N} \lp[
\mypmatrix{ 1 & \sqrt{N-1} \cr \sqrt{N-1} & N-1} 
+ \mypmatrix{ N-2 & 0 \cr 0 & 0} \rp ]
\nonumber\\
&=& \frac{\pi(N-1)}{N} I 
+ \frac{\pi\sqrt{N-1}}{N} \sigma_x,
\ee
\noindent
where $\ket{-_j}$ is the natural generalization of Eq.~(\ref{eq:ket-j}),
and $\sigma_x$ is the Pauli $x$ operator. The evolution is given
by
\be
\ket{\phi_j(t)} =& e^{-i\pi t (N-1)/N} 
\bigg(&\cos\lp(\frac{\pi t\sqrt{N-1}}{N}\rp) \ket{+} 
\\\nonumber
&& -
i\sin\lp( \frac{\pi t\sqrt{N-1}}{N}\rp) \ket{-_j} \bigg)
\ee
\noindent
and at time $T= N/(\pi \sqrt{N-1}) \arccos 1/\sqrt{N}$ we end up (ignoring
the global phase) in one of the states
\be
\ket{\phi_j(T)} \propto \frac{1}{\sqrt{N}} \ket{+} -
i \sqrt{\frac{N-1}{N}} \ket{-_j} = U \ket{j},
\ee
\noindent
where $U=-i I + (1+i) \ket{+}\bra{+}$. At this time all $N$ states become
mutually orthogonal, and therefore the $N$ different Hamiltonian oracles
can be perfectly distinguished.

In the protocol of Farhi and Gutmann \cite{FG98}, they used oracles of the
form $H_j = E\ket{j}\bra{j}$, and therefore to compare the results we need
to set $E=\pi$. With our notation their total Hamiltonian is given by
\be
H_{total} = H_j + H' = \pi \ket{j}\bra{j} + \pi \ket{+}\bra{+}
\label{eq:FG98}
\ee
and after a time of exactly $T=\sqrt{N}/2$ this Hamiltonian will evolve the
state $\ket{+}$ into the state $\ket{j}$. This is marginally slower than
the optimal time found above which can be expanded as $\sqrt{N}/2
-1/\pi+O(1/\sqrt{N})$. For $N=2$ the difference is exactly given by a
factor of $\sqrt{2}$ as pointed out in Ref.~\cite{CPR99}.

The practical difference between the two protocols is, of course,
insignificant. Nevertheless, it is interesting to see how it arises, as
generalizations of this trick will be useful later. In the relevant two
dimensional subspace for a given $j$, we can study the state on
the Block sphere, where we take the north pole to be the initial state
$\ket{+}$. The traditional goal is to evolve to the state $\ket{j}$ located
near the south pole, and the Hamiltonian of Eq.~(\ref{eq:FG98}) follows the
obvious path that connects them. However, the ability to add in a
Hamiltonian of arbitrary strength proportional to $\ket{+}\bra{+}$ is
equivalent to being able to do arbitrarily fast rotations around the
vertical axis. Therefore, the set of points on a circle of constant
latitude should all be regarded as a single point, and the optimal protocol
involves choosing at each time the correct longitude so that the evolution
southwards is greatest. In particular, the protocol need not arrive at
$\ket{j}$ but may end at any of the other points of similar latitude, which
the optimal protocol does.

The Farhi and Gutmann protocol \cite{FG98} does achieve a more general
goal: mainly given an oracle Hamiltonian $H=\pi \ket{m}\bra{m}$, where the
marked state $\ket{m}$ is arbitrary, evolve into the marked state (in a
time that depends only on the overlap of the marked with the initial
state). Our protocol essentially preassumes that the marked state is always
a computational basis state. However, if our goal is to identify the
Hamiltonian, then producing a copy of the marked state is only useful if
the set of possible marked states is orthogonal, in which case we may
assume that they belong to the computational basis.

\section{\label{sec:oi}Oracle Interrogation}

We now turn our attention to the Hamiltonian oracle version of Oracle
Interrogation \cite{vD98}, where the oracle has a $n$-bit string which can
be queried one bit at a time, and the goal is to output the complete
$n$-bit string. The problem is important as it serves as an upper bound on
all problems where the goal is to output some function of the $n$-bit
string.

We shall also briefly examine the \XOR problem, where the goal is
simply to output the \XOR of the above $n$ bits. In both the discrete and
continuous oracle setting, this problem is nearly as hard as outputting the
entire $n$-bit string.

\subsection{Problem definition}

Fix an integer $n\geq 1$, let $N=2^n$, and define
\be
\A &=& \vspan\{\ket{x}\text{ for }x\in\{0,1\}^n\},\\
\M &=& \vspan\{\ket{j,k}\text{ for }j\in\{0\}\cup[n],k\in\{0,1\}\}.
\ee

\noindent
We also introduce the final output spaces as $\M'=\A$ for oracle
interrogation and $\M'_\XOR = \vspan\{\ket{0},\ket{1}\}$ for the \XOR problem.
On these spaces define
\be
\ket{\psi_0} &=& \frac{1}{\sqrt{N}}\sum_{x}\ket{x},\\
O &=& \sum_{x,j,k} (-1)^{x_j+k}
\ket{x}\bra{x}_\A \otimes \ket{j,k}\bra{j,k}_\M,\\
H &=& \frac{\pi}{2} O,\\
\Pi_x &=& N\ket{x}\bra{x}_\A\otimes \ket{x}\bra{x}_{\M'},\\
\Pi_{x,\XOR} &=& N\ket{x}\bra{x}_\A\otimes
\ket{\XOR(x)}\bra{\XOR(x)}_{\M_\XOR'},
\ee

\noindent
where $x$ ranges over the $n$-bit strings, $j=0,\dots,n$ and
$k\in\{0,1\}$. We use the notation $x_j$ to denote the $j$th bit of $x$,
and define $x_0=0$. We also use $\XOR(x)$ to denote the $\XOR$ of the $n$
bits of $x$. The answer verification operators for oracle interrogation are
$\{\Pi_x\text{ for }x\in\{0,1\}^n\}$ whereas for the \XOR problem they are
$\{\Pi_{x,\XOR}\text{ for }x\in\{0,1\}^n\}$, otherwise the problems are
identical.

Though the oracle $O$ may look somewhat peculiar, it can be thought of as
the regular oracle that applies a phase $(-1)^{x_j}$ to query state
$\ket{j}$, followed by the $\sigma_z$ Pauli operator on the last
qubit. Since this operation is entirely on the message side, and could
equally well be applied by Bob before or after the query, and therefore
offers no extra computational power. As usual, Bob can also request a null
query on the state $\ket{0}$.

In the continuous case, the extra bit $k$ can be thought of as an arrow of
time. For $k=0$ the oracle applies one Hamiltonian and for $k=1$ the oracle
applies minus the same Hamiltonian. It is not clear whether one of these
blocks is computationally equivalent to the complete Hamiltonian. This is
an interesting open question. Unfortunately, the symmetrization approach to
studying the oracle requires both blocks.

The normalization of $H$ is chosen so that $e^{-i H} = -i O$, and hence the
query time for the Hamiltonian oracle problem is a lower bound on the
number of queries for the discrete oracle problem. The normalization does
have the unfortunate property that at time $t=1/2$ one can solve the $n=1$
case exactly. However, the unitary $e^{-i H/2}$ is equivalent to performing
the identity for a hidden zero bit, and applying phases of $\pm i$ for the
hidden one bit, and these operations cannot simulate the standard one bit
query.

\subsection{Symmetrization}

The natural symmetry group of these oracle problems is $G=(\Z_2)^{n}\rtimes
S_n$, with a multiplication rule given by $(x',s')(x,s)=(x'\oplus s'(x), s'
s)$ where $s\in S_n$ and $x$ is an $n$-bit binary string. The action of $s$
on $x$ is given by permutation of the bits.

The group $G$ has a set of representations defined by
\be
R_\A(x,s)\ket{y}_\A &=& \ket{x\oplus s(y)}_\A,\\
R_\M(x,s)\ket{j,k}_\M &=& \ket{s(j),x_{s(j)}\oplus k}_\M,\\
R_{\M'}(x,s)\ket{y}_{\M'} &=& \ket{x\oplus s(y)}_{\M'},\\
R_{\M'_\XOR}(x,s)\ket{k}_{\M'_\XOR} &=& \ket{\XOR(x)\oplus k}_{\M'_\XOR}.
\ee
With these definitions, both oracle problems are compatible with $G$.

We begin the symmetrization by describing the most general positive operator
$\rho$ on $\A$ that is $G$-invariant. We shall be working in the Hadamard
basis for $\A$ defined by
\be
\ket{\tilde x}_\A = H^{\otimes n}\ket{x}_\A,
\ee
\noindent
where $H$ is the qubit Hadamard operator. In this basis, the representation
$R_\A$ acts by
\be
R_\A(x,s) \ket{\tilde y} = (-1)^{|x\oplus s(y)|} \ket{\widetilde{s(y)}},
\ee
\noindent
where $|x|$ denotes the Hamming weight. Invariance under $G$ implies that
$\bra{\tilde x}\rho\ket{\tilde y}=\bra{\tilde x}R_\A(g^{-1})\rho
R_\A(g)\ket{\tilde y}$. Using elements of the form $g=(x',1)$ we see that
$\bra{\tilde x}\rho\ket{\tilde y}=0$ for $x\neq y$. Furthermore, using
$g=(1,s)$ we see that $\bra{\tilde x}\rho\ket{\tilde x}$ depends only on
the Hamming weight of $x$. We can therefore write the most general
$G$-invariant $\rho$ as
\be
\rho = \sum_{j=0}^n a_j^2 \lp(\frac{1}{\mypmatrix{n\cr j}}
\sum_{|x|=j}\ket{\tilde x}\bra{\tilde x} \rp).
\label{eq:irreps}
\ee
\noindent
The normalization is chosen so that $\Tr[\rho]=1$ implies $\sum_j a_j^2=1$
and therefore the vector $(a_0,\dots,a_n)$ is a point on the unit sphere
$S^n$ embedded in $\R^{n+1}$. Positivity of $\rho$ requires $a_j^2\geq 0$,
which in turn requires $a_j$ to be real. A unique set of $\rho$ matrices
can be generated by restricting to $a_j\geq 0$ for all $j$.

We now turn to the symmetrization of $\tilde \rho$. As $G$ acts by
permutation on $\M$, we can use the results of Section~\ref{sec:sigma}
which provide us with a decomposition of the most general $\tilde \rho$ as
\be
\tilde \rho = \sum_{j=0}^{n}\sum_{k=0}^{1} \sigma_{j,k}\otimes
\ket{j,k}\bra{j,k}_\M,
\ee
\noindent
where $\sigma_{0,0}$ and $\sigma_{0,1}$ must be $G$-invariant and the
restriction on the remaining $\sigma$ matrices is discussed below. We expand
\be
\sigma_{0,k} = \sum_{j=0}^n \zeta_{j,k} \lp(\frac{1}{\mypmatrix{n\cr j}}
\sum_{|x|=j}\ket{\tilde x}\bra{\tilde x} \rp).
\ee
\noindent
As the evolution will only depend on the sum $\sigma_{0,0}+\sigma_{0,1}$
(i.e., they both correspond to null queries), it will be convenient to
define $\zeta_j = \zeta_{j,0} + \zeta_{j,1}$, which are required to be
non-negative.

The remaining matrices are all related to each other by $\sigma_{j,k}=
R_A(g)\sigma_{1,0}R_A(g^{-1})$ for any $g\in G$ such that
$R_\M(g)\ket{1,0}_\M=\ket{j,k}_\M$. In particular, $\sigma_{1,0}$ must be
invariant under the subgroup $H$ of $G$ that leaves $\ket{1,0}_\M$
invariant. We can write $H=(Z_2)^{n-1}\rtimes S_{n-1}$.

From Eq.~(\ref{eq:irreps}) we see that $\A$ decomposes into $n+1$ irreps of
$G$ given by $\vspan\{\ket{\tilde x}\text{ for }|x|=j\}$ for
$j=0,\dots,n$. The $j=0$ and $j=n$ irreps are both one dimensional and
therefore will also be irreps of $H$. Under the restriction to the subgroup
$H$, the other irreps each split into two. An irrep of vectors with Hamming
weight $j$ will split into the vectors that have a zero in the first slot
(which will be an irrep of $H$ consisting of Hamming weight $j$ vectors),
and the vectors that have a one in the first slot (which will be an irrep
of $H$ consisting of Hamming weight $j-1$ vectors). In total, we end up
with two copies of each of the $n$ irreps. Each pair of irreps can share
off diagonal elements but otherwise the matrix must be block
diagonal. Therefore the most general $H$-invariant operator on $\A$ has the
form
\be
\sigma_{1,0} =  \sum_{j=0}^{n-1} &\Bigg[&
\mypmatrix{\ket{\tilde 0}&\ket{\tilde 1}} 
\mypmatrix{\alpha_j & \beta_j\cr \beta_j^* & \gamma_j}
\mypmatrix{\bra{\tilde 0}\cr\bra{\tilde 1}}
\\\nonumber
&&~\otimes \frac{1}{2n\mypmatrix{n-1\cr j}}
\sum_{\substack{z\in\{0,1\}^{n-1}\cr |z|=j}} \ket{\tilde z}\bra{\tilde z}
\Bigg],
\ee
\noindent
where we have decomposed $\A$ into the first qubit and the remaining $n-1$
qubits. The notation means that, for instance, $\ket{\tilde
0}\otimes\ket{\tilde z}=\ket{\tilde x}$ where $x$ is the $n$-bit string
obtained by concatenating $0$ and $z$. Positivity of $\sigma_{1,0}$ is
equivalent to $\alpha_j\geq 0$, $\gamma_j\geq 0$ and $\alpha_j \gamma_j\geq
|\beta_j|^2$ for every $j$.

Note that the above form for an
$H$-invariant $\sigma_{1,0}$ could also be obtained directly by noting that
$H$-invariance implies that $\bra{x}\sigma_{1,0}\ket{y}$ can depend only on
the Hamming weight of the last $n-1$ bits of $x\oplus y$.

The normalizations above have been chosen in order to simplify the
equation $\rho=\Tr_\M[\tilde \rho]$ which is now equivalent to
\be
a_j^2 =
\begin{cases}
\zeta_0 + \alpha_0 & j=0,\\
\zeta_j + \alpha_j + \gamma_{j-1}
& 0<j<n,\\
\zeta_n + \gamma_{n-1} & j=n.
\end{cases}
\label{eq:cons}
\ee

\subsection{Boundary conditions}

The initial condition is simply given by $a_0=1$ and $a_j=0$ for $j>0$.
For the final probabilities of success, we note that after symmetrization,
the probability of correctly outputting $x$ or $\XOR(x)$ is independent of
the hidden string $x$, therefore we can replace the final measurements by
\be
\Pi &=& \sum_x \ket{x}\bra{x}_\A\otimes\ket{x}\bra{x}_{\M'}\\
\Pi_{\XOR} &=& \sum_x \ket{x}\bra{x}_\A\otimes
\ket{\XOR(x)}\bra{\XOR(x)}_{\M'_\XOR}
\ee
\noindent
for the oracle interrogation and $\XOR$ problems respectively. The final
step is then a standard state discrimination problem dependent only on
$\rho(T)$.

For oracle interrogation, $\rho(T)$ is proportional to the Gram matrix of
the states to be distinguished. Since it is diagonal in the Hadamard basis,
$\sqrt{\rho(T)}$ is also diagonal in the Hadamard basis and hence its
diagonal elements in the computational basis are all equal. Just as in the
Grover search case above, this implies \cite{BKM97,SKI98,Moc05} that the
optimal measurement is the pretty good measurement and the success
probability is given by
\be
P_{win}(\rho(T))&=&\frac{1}{N}\lp(\Tr\sqrt{\rho(T)}\rp)^2 \\\nonumber
&=& \frac{1}{N} \lp( \sum_j \sqrt{\mypmatrix{n\cr j}} a_j(T)\rp)^2 
= \lp( \vec a(T) \cdot \vec a_f \rp)^2.
\ee
\noindent
where $\vec a(T) =(a_0(T),\dots,a_n(T))$ which involves the components of
$\rho(T)$. The target vector $\vec a_f$ with components
\be
(\vec a_f)_j = \sqrt{\frac{1}{N}\mypmatrix{n\cr j}}
\ee
has unit length, and so a zero error solution requires $\vec a(T)=\vec
a_f$.

The last step in the \XOR problem involves the state discrimination
of two mixed states. As Bob has the purification of $\rho(T)$, we can write
the joint state as $\sqrt{\rho(T)}\otimes I\ket{\Phi}$ where
$\ket{\Phi}=\sum_x \ket{x}_\A\otimes\ket{x}_{\M'}$. The two states to
discriminate are therefore given by
\be
\eta_k &=& \Tr_\A \lp[ P_k \otimes I 
\lp(\sqrt{\rho(T)}\otimes I\rp) \ket{\Phi}
\bra{\Phi} \lp(\sqrt{\rho(T)}\otimes I\rp) \rp]
\nonumber\\ 
&=& \lp( \sqrt{\rho(T)} P_k \sqrt{\rho(T)} \rp)^T,
\ee
\noindent
where $k\in\{0,1\}$, $P_k$ is the projector onto states $\ket{x}$ with
$\XOR(x)=k$, and the transpose is taken in the computational basis. The
normalization is set to $\Tr[\eta_k]=1/2$ which is the a priori
probability. Now we can use the result of Helstrom for two-state
discrimination \cite{Hel76}, so that
\be
P_{win}(\rho(T)) &=& \frac{1}{2} \Tr\lp( 
\eta_0 + \eta_1 + \lp|\eta_0 - \eta_1 \rp| \rp)
\\\nonumber
&=& \frac{1}{2}+ \frac{1}{2}
\Tr\lp| \sqrt{\rho(T)} \lp (P_0-P_1 \rp) \sqrt{\rho(T)}\rp|
\\\nonumber
&=& \frac{1}{2} + \frac{1}{2} \sum_{j=0}^n a_j(T) a_{n-j}(T),
\ee 
\noindent
where in the last step we use the fact that $P_0-P_1=\sigma_x^{\otimes n}$
in the Hadamard basis, and so the matrix inside the absolute value is block
diagonal with blocks pairing $\ket{\tilde x}$ and $\ket{\widetilde{x\oplus
1\cdots 1}}$.

From the above discussion we can see that the zero error \XOR final states
satisfy $a_j = a_{n-j}$ for all $j$. Furthermore if $a_j=0$ for $j\geq n/2$
then $P_{win}\leq 1/2$ for both the \XOR and oracle interrogation problems.
In fact, we have for both problems
\be
P_{win} \leq \frac{1}{2} + \sqrt{\sum_{j=\lp \lfloor \frac{n}{2}
\rp \rfloor}^n a_j^2(T)}.
\label{eq:winbd}
\ee

\subsection{Discrete oracle dynamics}

We shall only sketch the discrete oracle case here for comparison. From
Eq.~(\ref{eq:O}) we get the dynamics
\beq
\rho(t+1) = \sigma_{0,0}(t) + \sigma_{0,1}(t) 
+ 2 n\lp[ O_{1,0} \sigma_{1,0}(t) O_{1,0}^{-1}\rp]_{sym},
\eeq
\noindent
where $[~]_{sym}$ refers to the projection to the $G$-invariant subspace. The
operator $O_{1,0}$ is the unitary realized when state $\ket{1,0}_\M$ is
queried, and has the effect exchanging $\alpha_i\leftrightarrow \gamma_i$
and $\beta\leftrightarrow\beta^*$, leading to the equations
\beq
a_j^2(t+1) =
\begin{cases}
\zeta_0(t) + \gamma_0(t) & j=0,\\
\zeta_j(t) + \gamma_j(t) + \alpha_{j-1}(t)
& 0<j<n,\\
\zeta_n(t) + \alpha_{n-1}(t) & j=n.
\end{cases}
\eeq
\noindent
In combination with the constraint Eq.~(\ref{eq:cons}), one can see that at
every step we can split $a_j^2(t)$ into three pieces: one which will get
added into $a_{j+1}^2(t+1)$, one which will be added into $a_{j-1}^2(t+1)$
and one which will remain in $a_{j}^2(t+1)$. Inductively, we can prove that
the set of achievable vectors after $t$ queries satisfy $a_j=0$ for $j>t$
but otherwise need only satisfy the normalization constraint $\sum_j
a_j^2=1$.

In particular, this proves that for the \XOR problem $P_{win}=1/2$ for
$T<n/2$ whereas $P_{win}=1$ for $T=\lceil n/2 \rceil$, which is achieved as
follows: for $n$ even $a_{n/2}^2=1$ and the rest zero, for $n$ odd
$a_{(n\pm 1)/2}^2=1/2$ and the rest zero.

For oracle interrogation we see that an exact solution requires $T=n$.
However, since most of the amplitude of the final vector $\vec a_f$ is
contained in the indices $a_{n/2\pm O(\sqrt{n})}$ the problem can be solved to
high accuracy by only correctly adjusting these components. This requires a
query time $T=n/2+O(\sqrt{n})$ reproducing the result of van Dam
\cite{vD98}.

\subsection{Continuous oracle dynamics}

From Eq.~(\ref{eq:H}) we get the dynamics
\be
\frac{d \rho(t)}{dt} = -2n i \lp[ H_{1,0} \sigma_{1,0}(t) - \sigma_{1,0}(t)
H_{1,0} \rp]_{sym},
\ee
\noindent
where as before $[~]_{sym}$ is the projection onto the symmetric subspace
and $H_{1,0} = \frac{\pi}{2} \sum_x (-1)^{x_1} \ket{x}\bra{x}$. In each of
the $2\times 2$ blocks comprising $\sigma_{1,0}$, $H_{1,0}$ is proportional
to the Pauli $\sigma_x$ operator (as the blocks are in the Hadamard basis)
and hence each block leads to a calculation of the form
\be
&&\mypmatrix{ 0  & 1 \cr 1 & 0}
\mypmatrix{\alpha_j & \beta_j\cr \beta_j^* & \gamma_j}
-
\mypmatrix{\alpha_j & \beta_j\cr \beta_j^* & \gamma_j}
\mypmatrix{ 0  & 1 \cr 1 & 0}
\\\nonumber
&&
\qquad\qquad\qquad\qquad\qquad
= \mypmatrix{-2i\Imag[\beta_j] & \gamma_j-\alpha_j \cr 
\alpha_j - \gamma_j & 2i\Imag[\beta_j]}
\ee
\noindent
which leads to the differential equations
\beq
\frac{d a^2_j(t)}{dt} 
= -\pi 
\begin{cases}
\Imag \beta_0(t) & j=0,\\
\Imag \beta_j(t) - \Imag \beta_{j-1}(t) 
& 0<j<n,\\
-\Imag \beta_{n-1} & j=n.
\end{cases}
\eeq
\noindent
Now we apply the constraints from the positivity of $\sigma_{1,0}$ which
imply $|\Imag\beta_j(t)| \leq \sqrt{\alpha_j(t) \gamma_j(t)}$. From
Eq.~(\ref{eq:cons}) we also have $a_j^2 \leq \alpha_j + \gamma_{j-1}$ (with
$\alpha_n = \gamma_{-1} = 0$). We can therefore write at every time
\be
\Imag \beta_j(t) = b_j(t) c_{j+1}(t) a_j(t) a_{j+1}(t),
\ee
\noindent
where the new parameters represent Bob's degrees of freedom but must be
consistent with the constraint $b_j^2(t) + c_j^2(t) \leq 1$ for
$j=0,\dots,n$. Canceling a factor of $a_j(t)$ we obtain
\be
\frac{d}{dt} \vec a(t) = M(t) \vec a(t),
\label{eq:rot}
\ee
\noindent
where $M(t)$ is the $(n+1)\times(n+1)$ real antisymmetric (as required by
probability conservation) matrix which is zero everywhere except the
entries one-off from the diagonal
\be
M(t)_{j,j+1} = -M(t)_{j+1,j} = -\frac{\pi}{2} b_j(t) c_{j+1}(t) 
\label{eq:M}
\ee
\noindent
for $j=0,\dots,n-1$. An extra factor of $1/2$ appears in the above equation
from the relation $\frac{d a^2}{dt}=2a\frac{d a}{dt}$.

Note that in the transition to Eq.~(\ref{eq:rot}) we canceled factors of
$a_j(t)$ which potentially could be zero. All this implies is that the
derivative of $a_j(t)$ need not satisfy the above equation when
$a_j=0$. However, this is a set of measure zero, and a continuous evolution
of $\vec a$ will require that the above equation be satisfied at all times.

Let us rehash the current state of the problem. The vector $\vec{a}(t)$
indicates the state of the system (and hence Bob's knowledge of the hidden
string) at a given time. Bob can affect this parameter by controlling the
matrix $M(t)$ which he can modify at any time. The matrix $M(t)$ must have
the form given by Eq.~(\ref{eq:M}) with $b_j^2(t) + c_j^2(t) \leq 1$ but
otherwise can be chosen arbitrarily. Bob must choose the parameters
$\{b_j(t),c_j(t)\}$ to evolve from the initial condition of
$\vec{a}(0)=(1,0,\dots,0)$ in order to maximize $\vec{a}(T)\cdot \vec{a}_f$
at some final time $T$ and with high probability solve the oracle
interrogation problem.  A similar end criterion was formulated above for
the \XOR problem.

Unfortunately, finding such an optimal evolution is still a difficult
problem. We shall find below the optimal strategies for zero-error oracle
interrogation for $n=1$ and $n=2$. The latter case is obtained by studying
the geodesics of $S^2$ with a Riemannian metric. For $n>2$ the metrics
appear to be of Finsler type, and therefore beyond the scope of this
paper. Nevertheless, we shall also prove a simple lower bound that will
apply both to the oracle interrogation and \XOR problems and will apply to
bounded error solutions as well.

\subsection{The $n=1$ case}

For $n=1$ the differential equation reads
\be
\frac{d}{dt} \mypmatrix{a_0(t)\cr a_1(t)} = 
-\frac{\pi}{2} b_0(t) c_1(t) 
\mypmatrix{0 & 1 \cr -1 & 0} \mypmatrix{a_0(t)\cr a_1(t)}
\ee
\noindent
with constraints $b_0^2\leq 1$ and $c_1^2\leq 1$ (note that $c_0$ and $b_1$
do not appear anywhere in the equation). The initial condition is
$\vec{a}(0)=(1,0)$ and the final vector for zero error oracle interrogation
is $\vec{a}(T) = \vec{a}_f = (1,1)/\sqrt{2}$.

The optimal algorithm is to choose $b_0(t) = c_1(t) = 1$ at all times, in
which case we obtain the evolution
\be
\mypmatrix{a_0(t)\cr a_1(t)} = 
\mypmatrix{\cos \pi t/2 \cr \sin \pi t/2}.
\ee
The minimum time required to arrive at the zero error final point is
$T=1/2$.

\subsection{The $n=2$ case}
For $n=2$ the differential equation reads
\beq
\frac{d}{dt} \mypmatrix{a_0(t)\cr a_1(t)\cr a_2(t)} = 
\frac{\pi}{2}  
\mypmatrix{0 & -w_1(t) & 0 \cr w_1(t) & 0 & -w_2(t) \cr 0 & w_2(t) & 0 } 
\mypmatrix{a_0(t)\cr a_1(t)\cr a_2(t)}
\eeq
\noindent
with $w_1=-b_0 c_1$ and $w_2=-b_1 c_2$, which are constrained by
$w_1^2+w_2^2\leq 1$.

If we position the unit sphere so that the vector $(0,1,0)$ corresponds
with the north pole, then effectively, Bob can perform any rotation around
an axis that lies on the equator and at a speed less than or equal to
$\pi/2$ radians per unit time. Rotations around other axes can only be
generated as composite rotations.

Thus far we have restricted ourselves to vectors $\vec a$ from the
intersection of the non-negative cone with the unit sphere. However we can
now lift the restriction and allow vectors from the entire unit sphere. The
only consequence of this is that we must identify points that differ by
changes of sign, as the real state $\rho(t)$ depends only on $a_i^2(t)$. We
now have to consider two possible starting points and eight possible
zero-error ending points. The symmetry (reflections north-south and
east-west) reduces the set of inequivalent pairs to only two: starting from
$(1,0,0)$ and ending at either $(1,\sqrt{2},1)/2$ or
$(-1,\sqrt{2},1)/2$. Note that the paths that connect to the latter point
would still be allowed under the restriction to the non-negative cone, but
would have required a ``bounce'' on a boundary.

We shall now reformulate the problem in the language of differential
geometry, where the sphere will acquire a non-round metric constructed so
that the shortest distance between two points is equal to the minimum query
time that is needed to evolve from one point to the other. The notation
used below will follow the conventions adopted in general relativity.

It will be convenient to work in polar coordinates
\be
a_0 &=& \sin\theta\cos\phi,\\
a_1 &=& \cos\theta,\\
a_2 &=& \sin\theta\sin\phi,
\ee
\noindent
where the initial condition is now $\theta=\pi/2$, $\phi=0$ and the
final points are $\theta=\pi/4$, $\phi=\pi/2\pm \pi/4$. Associated to this
basis we have the coordinate (unnormalized) basis for the tangent space
\be
e_{\theta} &=& (\cos\theta\cos\phi, -\sin\theta, 
\cos\theta\sin\phi),\\
e_{\phi} &=& (-\sin\theta\sin\phi, 0, \sin\theta\cos\phi),
\ee
At any given time, the set of possible velocity vectors depends on the
current position and the Bob controlled parameters $w_1$, $w_2$ and is
given by
\beq
\frac{\pi}{2}  
\mypmatrix{0 & -w_1 & 0 \cr w_1 & 0 & -w_2 \cr 0 & w_2 & 0 } 
\mypmatrix{a_0\cr a_1\cr a_2}
=  \frac{\pi}{2} \lp ( w_\theta e_\theta + \frac{w_\phi}{\tan \theta}
e_\phi \rp),
\eeq
\noindent
where we introduced
\be
w_\theta &=& w_2 \sin\phi - w_1 \cos\phi,\\
w_\phi &=& w_2 \cos\phi + w_1 \sin\phi.
\ee
The constraint $w_1^2+w_2^2\leq1$ is equivalent to $w_\theta^2 + w_\phi^2\leq
1$. It is always optimal for Bob to choose the magnitude of the velocity to
be as large as possible consistent with the chosen direction, and hence the
inequality constraint will always be saturated. This produces a set of
velocity vectors that correspond to unit velocity. The same set can be
generated by the metric
\be
ds^2 = \frac{4}{\pi^2}\lp( d\theta^2 + \tan^2 \theta d\phi^2 \rp) 
\ee
\noindent
and therefore the distance assigned to a curve by this metric will be equal
to the time it would take Bob to evolve the system through that curve. One
is now left with the problem of finding curves of minimal distance on the
surface with the above metric.

Strictly speaking the metric is ill defined on the equator, where our
initial point lies. One can instead study curves that begin at
$\theta=\pi/2-\epsilon$ and $\phi=0$ and then bound the distance of these
points to the equator. The resulting total distance in the limit
$\epsilon\rightarrow 0$, however, will be the same as will be derived below
by ignoring the divergence at the equator.

We can also describe the metric by its non-zero components
$g_{\theta\theta} = \frac{4}{\pi^2}$ and $g_{\theta\theta} =
\frac{4}{\pi^2}\tan^2\theta$. The Christoffel symbols are 
defined by
\be
\Gamma^\lambda_{\mu\nu} = \frac{1}{2}g^{\lambda \delta} 
\lp(g_{\mu\delta,\nu} + g_{\delta\nu,\mu} - g_{\mu\nu,\delta}\rp) 
\ee
\noindent
and therefore the non-zero symbols for our metric are given by
\be
\Gamma^\theta_{\phi\phi} &=& -\frac{1}{2}\frac{\partial}{\partial\theta} 
\tan^2\theta = - \frac{\sin\theta}{\cos^3\theta},\\
\Gamma^\phi_{\theta\phi}=\Gamma^\phi_{\phi\theta} &=& 
\frac{1}{2\tan^2\theta}\frac{\partial}{\partial\theta} 
\tan^2\theta = \frac{1}{\sin\theta\cos\theta}.\qquad
\ee
The geodesic equation is
\be
\frac{d v^\lambda}{dt} = - \Gamma^{\lambda}_{\mu\nu} v^{\mu} v^{\nu},
\ee
\noindent
where $v$ is the velocity vector. Using dots for time derivatives the
geodesic differential equations for our metric can be written as
\be
\ddot \theta &=& \frac{\sin\theta}{\cos^3\theta}\, \dot \phi^2,\\
\ddot \phi &=& \frac{-2}{\sin\theta\cos\theta} \,\dot \theta\, \dot \phi.
\ee
The second equation is solved by
\be
\dot \phi = \pm \frac{\pi}{2} \frac{\tan\theta_0}{\tan^2\theta},
\label{eq:dphidt}
\ee
\noindent
where $\theta_0$ is an arbitrary parameter whose form will become clear in
a moment. The same equation can also be obtained directly by the variation
of the action with respect to $\phi$.  The geodesic equation also implies
the conservation of the total speed, which we normalize to one
\be
1 = g_{\theta\theta}\dot\theta^2 + g_{\phi\phi}\dot\phi^2.
\ee
We can now combine the two previous equations to obtain a differential
equation for $\theta$
\be
\dot \theta = \pm \frac{\pi}{2}
\sqrt{1 - \frac{\tan^2\theta_0}{\tan^2\theta}},
\ee
\noindent
where the meaning of $\theta_0$ becomes clear: it defines the maximum
height of the geodesic curve. The differential equation is solved by
\be
\frac{\pi}{2} t &=& \pm 
\int \frac{\cos(\theta_0) \sin\theta d\theta}
{\sqrt{\cos^2\theta_0 \sin^2 \theta - \sin^2\theta_0\cos^2\theta}}
\nonumber\\
&=&
\mp \int \frac{\cos \theta_0 d(\cos \theta)}
{\sqrt{\cos^2\theta_0 - \cos^2\theta}}
\nonumber\\
&=& \cos \theta_0 \arcsin\lp( \frac{\cos \theta}{\cos \theta_0} \rp),
\ee
\noindent
where in the last step we have chosen our constant and sign so that $t=0$
corresponds to the initial condition of $\theta=\pi/2$, and as time
increases we move north.

Now we turn to the differential equation for $\phi$ which can be obtained
by substituting the above solution into Eq.~(\ref{eq:dphidt})
\be
\dot \phi &=& \pm\frac{\pi \tan \theta_0}{2}\lp( \frac{1}{1 - \cos^2 \theta} 
- 1 \rp)
\\\nonumber &=&
\pm\frac{\pi \tan \theta_0}{2}\lp( \frac{1}{1 - \cos^2 \theta_0 
\sin^2 \frac{\pi t}{2\cos\theta_0}} - 1 \rp). 
\ee
Using the derivative
\be
\frac{d}{ds}\arctan\lp(\sin\theta_0 \tan s\rp)
&=& \frac{\sin \theta_0}{\cos^2 s}
\lp[1+ \sin^2\theta_0 \tan^2 s\rp]^{-1}
\nonumber\\
&=& \frac{\sin \theta_0}{1-\cos^2\theta_0\sin^2 s}
\ee
we obtain
\be
\label{eq:phi}
\phi &=& - \frac{\pi t \tan \theta_0}{2} +
\arctan \lp( \sin\theta_0 \tan\frac{\pi t}{2\cos\theta_0}\rp) 
\\\nonumber
&=& - \sin\theta_0 \arcsin\lp(\frac{\cos \theta}{\cos \theta_0}\rp)
+ \arctan\lp( \frac{\sin \theta_0 \frac{\cos \theta}{\cos \theta_0}}
{\sqrt{1- \frac{\cos^2 \theta}{\cos^2 \theta_0}}}
\rp)
\ee
\noindent
with a choice of the additive constant and sign so that $\phi=0$ at $t=0$,
and $\phi$ increases with time. Unfortunately, solving for the constant
$\theta_0$ seems to require solving a transcendental equation, and
therefore the calculation needs to be completed numerically.

Of course there are many geodesics that connect the points that we are
interested in. Before proceeding with a numerical solution, we must ensure
that we are examining the shortest geodesic.

The geodesics all start at the equator, rise up to some height $\cos
\theta_0$, and then fall back again to the equator so that the curve is
symmetric around the apex. During the transition from $\theta=\pi/2$ to
$\theta=\theta_0$ we effect the following increases:
\be
\Delta t &=& \cos\theta_0,\\
\Delta \phi &=& \frac{\pi}{2} (1-\sin\theta_0).
\label{eq:deltaphi}
\ee
\noindent
Also note that if we remove the $\sin\theta_0$ factor from inside the
$\arctan$ we increase the right-hand side of Eq.~(\ref{eq:phi}). Without
that factor however, the $\arctan$ is equivalent to and $\arcsin$ and so we
have
\be
\phi \leq (1-\sin\theta_0) \arcsin\lp(\frac{\cos \theta}{\cos \theta_0}\rp).
\ee
We learn two things from the above observations. First, we learn that on the
way up, $\phi\leq(1-1/\sqrt{2})\pi/2\leq \pi/4$ at $\theta=\pi/4$ so that we
must pass the apex at least once before arriving at the zero-error
solution. Second, we need a solution with $\theta_0\leq\pi/4$ and hence the
time to climb to the apex and return to the equator is at least
$2/\sqrt{2}>1$ which is more time than it takes to query the two bits
separately. Therefore the optimal solution must rise to the apex once, and
arrive at either $\phi=\pi/4$ or $\phi=3\pi/4$ on the way down. The time of
arrival for such a trip is
\be
T = 2\cos \theta_0\lp(1  - \frac{1}{\pi} \arcsin\lp(\frac{\cos
\theta}{\cos \theta_0}\rp)\rp),
\ee
which increases as $\cos \theta_0$ gets larger. Since a geodesic to
$\phi=3\pi/4$ will require a larger $\cos \theta_0$ than one to
$\phi=\pi/4$ we have proven that the shortest path to a zero error point
arrives at $\phi=\pi/4$ after crossing through the apex exactly once.  The
total increase in $\phi$ over such a path is given by twice the right hand
side of Eq.~(\ref{eq:deltaphi}) minus the right hand side of
Eq.~(\ref{eq:phi}). Substituting into this equation $\theta=\pi/4$ and
$\phi=\pi/4$ we can numerically solve for $\cos \theta_0 \simeq
0.7477$. Using this value in the above equation we find that the query time
needed to exactly solve the $n=2$ case of oracle interrogation is
\be
T \simeq 0.9052.
\ee
That is, only about $90\%$ of the time it would require to query both bits
separately.

\subsection{\label{sec:low}Lower bound}

To conclude we shall prove a weak but fairly simple lower bound on the
query time needed to solve the Hamiltonian oracles for \XOR and oracle
interrogation even in the bounded error setting.

From the discrete case we learn that in general amplitude moves from the
variables $a_j$ with low values of $j$ to the ones with high values of
$j$. We also know that after only $t$ queries, the variables $a_j$ with
$j>t$ are zero. Though this no longer holds in the continuous case, it does
motivate the study of the variables
\be
A_j = \sqrt{\sum_{k=j}^n a_k^2 }.
\ee
From the dynamical of Eq.~(\ref{eq:rot}) we have for $j>0$
\be
\frac{d A_j^2(t)}{dt} &=& 2 \sum_{k=j}^n  a_k(t) \frac{d a_j(t)}{dt} 
\\\nonumber
&=& \pi b_{j-1}(t) c_j(t) a_j(t) a_{j-1}(t).
\ee
\noindent
Using $A_j\geq a_j$, $b_{j-1}\leq 1$ and $c_j\leq 1$ we obtain
\be
\frac{d A_j(t)}{dt}\leq \frac{\pi}{2} A_{j-1}(t).
\ee
\noindent
An inductive solution can be constructed because we know that for all time
$A_0(t)=1$ by conservation of probability, and the remaining initial
conditions are $A_j(0)=0$ for $j>0$. Therefore
\be
A_j(t) \leq \frac{1}{j!} \lp( \frac{\pi t}{2}\rp)^j.
\ee
\noindent
In particular, we know from Eq.~(\ref{eq:winbd}) that we can relate the
probability of success to the above variables by $P_{win}(T) \leq
\frac{1}{2} + A_{\lfloor n/2\rfloor}(T)$. Therefore, the query time 
needed to solve the Hamiltonian versions of \XOR and oracle interrogation
with bounded error is at least
\be
T &\geq& \frac{2}{\pi}\lp(\lp\lfloor \frac{n}{2}\rp\rfloor!\rp)^{1/\lfloor
\frac{n}{2}\rfloor} \lp|P_{win}(T)-\frac{1}{2} \rp|^{1/\lfloor
\frac{n}{2}\rfloor}
\nonumber\\
&\geq&
\frac{n}{\pi e} + \Omega(1) \simeq 0.117 n .
\ee

The bound is likely weak in the continuous case, and certainly weak as a
lower bound of the discrete case. Nevertheless, it captures the essential
$O(n)$ scaling. The main open question is: can similar continuous methods be
used to prove lower bounds for new problems?

\begin{acknowledgments}
Research at Perimeter Institute for Theoretical Physics is supported in
part by the Government of Canada through NSERC and by the Province of
Ontario through MRI. The author would like to thank Michael Nielsen for
providing an early copy of his latest manuscript, and for his hospitality
in Brisbane where this work was begun. Helpful discussions were provided by
Graeme Smith, Andrew Childs and Debbie Leung.
\end{acknowledgments}



\end{document}